\documentclass[fleqn,usenatbib]{mnras}

\usepackage{newtxtext,newtxmath,relsize}

\usepackage[T1]{fontenc}

\DeclareRobustCommand{\VAN}[3]{#2}
\let\VANthebibliography\thebibliography
\def\thebibliography{\DeclareRobustCommand{\VAN}[3]{##3}\VANthebibliography}

\usepackage{graphicx}	
\usepackage{amsmath}	
\usepackage{textcomp}

\newcommand{\be}{\begin{equation}}
\newcommand{\ee}{\end{equation}}
\newcommand{\psra}{\mbox{PSR J1939$+$2134}}
\newcommand{\psrb}{\mbox{PSR J0837$+$0610}}

\title[H-FISTA: Phase retrieval for pulsars]{H-FISTA: A hierarchical algorithm for phase retrieval with application to pulsar dynamic spectra}

\author[Os\l owski \&\ Walker]{
Stefan Os\l owski,$^{1}$\thanks{E-mail: stefan.oslowski@manlyastrophysics.org}
Mark A. Walker$^{1}$\\
$^{1}$Manly Astrophysics, 15/41-42 East Esplanade, Manly, NSW 2095, Australia}

\date{Accepted XXX. Received YYY; in original form ZZZ}

\pubyear{2015}

\begin{document}
\label{firstpage}
\pagerange{\pageref{firstpage}--\pageref{lastpage}}
\maketitle

\begin{abstract}
A pulsar dynamic spectrum is an inline digital hologram of the interstellar medium; it encodes information on the propagation paths by which signals have travelled from source to telescope. To decode the hologram it is necessary to ``retrieve'' the phases of the wavefield from intensity measurements, which directly gauge only the field modulus, by imposing additional constraints on the model. We present a new method for phase retrieval in the context of pulsar spectroscopy. Our method makes use of the Fast Iterative Shrinkage Thresholding Algorithm (FISTA) to obtain sparse models of the wavefield in a hierarchical approach with progressively increasing depth. Once the tail of the noise distribution is reached the hierarchy terminates with a final, unregularised optimisation. The result is a fully dense model of the complex wavefield that permits the discovery of faint signals by appropriate averaging. We illustrate the performance of our method on synthetic test cases and on real data. Our algorithm, which we call H-FISTA, is implemented in the Python programming language and is freely available.
\end{abstract}




\section{Introduction}
\message{The column width is: \the\columnwidth}

Although not the most numerous of astronomical radio sources, pulsars are individually amongst the most informative for studies of multipath propagation of radio waves through the interstellar medium (ISM) \citep[e.g.][]{1990ARA&A..28..561R}. Interference fringes between these various paths have high visibility, because of the small size of the pulsar radio emission region, and a single dynamic spectrum may contain a great deal of information on the structure and kinematics of the scattering material (free electrons, typically) along the line-of-sight to the source. Studying these propagation effects is important both for understanding the ISM itself and for characterising the associated signal delays --- which can be a significant source of systematic error in pulsar timing experiments \citep[e.g.][]{2004hpa..book.....L,Verbiest2020}.

Many studies of radio pulse propagation in the ISM have concentrated on statistical properties, such as the spectral and temporal widths of the signal autocorrelations, i.e., the scintillation timescale and bandwidth (or its inverse, the pulse broadening time). These quantities can be readily measured \citep[e.g.][]{1985ApJ...288..221C,1994MNRAS.269.1035G,2004ApJ...605..759B}, and theoretical predictions are available for a variety of models of the scattering material \citep[e.g.][]{1985MNRAS.214..519G,1999ApJ...517..299L}. A diverse collection of statistical properties, including measures that are not derived from pulsar observations, have been shown to be broadly consistent with a Kolmogorov spectrum of inhomogeneities in the ionised ISM over a very wide range of spatial scales \citep[][]{1995ApJ...443..209A}. However, the observational data on both pulsars and quasars sometimes show features that are inconsistent with Kolmogorov turbulence \citep[e.g.][]{1986ApJ...307L..27C,1987Natur.326..675F,1997MNRAS.287..739R,2010ApJ...708..232B,2017MNRAS.470.2659G}. These anomalies are not understood, but it is clear that they cannot be fully characterised by a small number of statistical measures. The high information content of individual pulsar dynamic spectra makes spectroscopy an attractive approach for investigating the cause of these anomalies, and to that end techniques for extracting the information are needed.

One such technique that has proved valuable is to study the power spectrum of the dynamic spectrum --- a quantity that is often referred to as the ``secondary spectrum''. Applying that method to high resolution, high signal-to-noise data \citet{2001ApJ...549L..97S} discovered that power is often concentrated along parabolic loci in the Fourier domain, with signal delay (conjugate to radio frequency) being proportional to the square of the signal doppler-shift (conjugate to the sample time). The preponderance of parabolic arcs in pulsar secondary spectra is now understood to be entirely a result of the scattering geometry --- see  \citet{2006ApJ...637..346C} and \citet{2004MNRAS.354...43W}. The material responsible for scattering the radiation is not distributed uniformly along the whole line-of-sight, but is tightly concentrated at one or more distances. Thus the scattering medium can be considered to be a collection of thin screens. And the visibility of the scintillation arcs is greatly enhanced if the scattering is significantly anisotropic.

\begin{figure*}
    \centering
    \includegraphics[width=\textwidth]{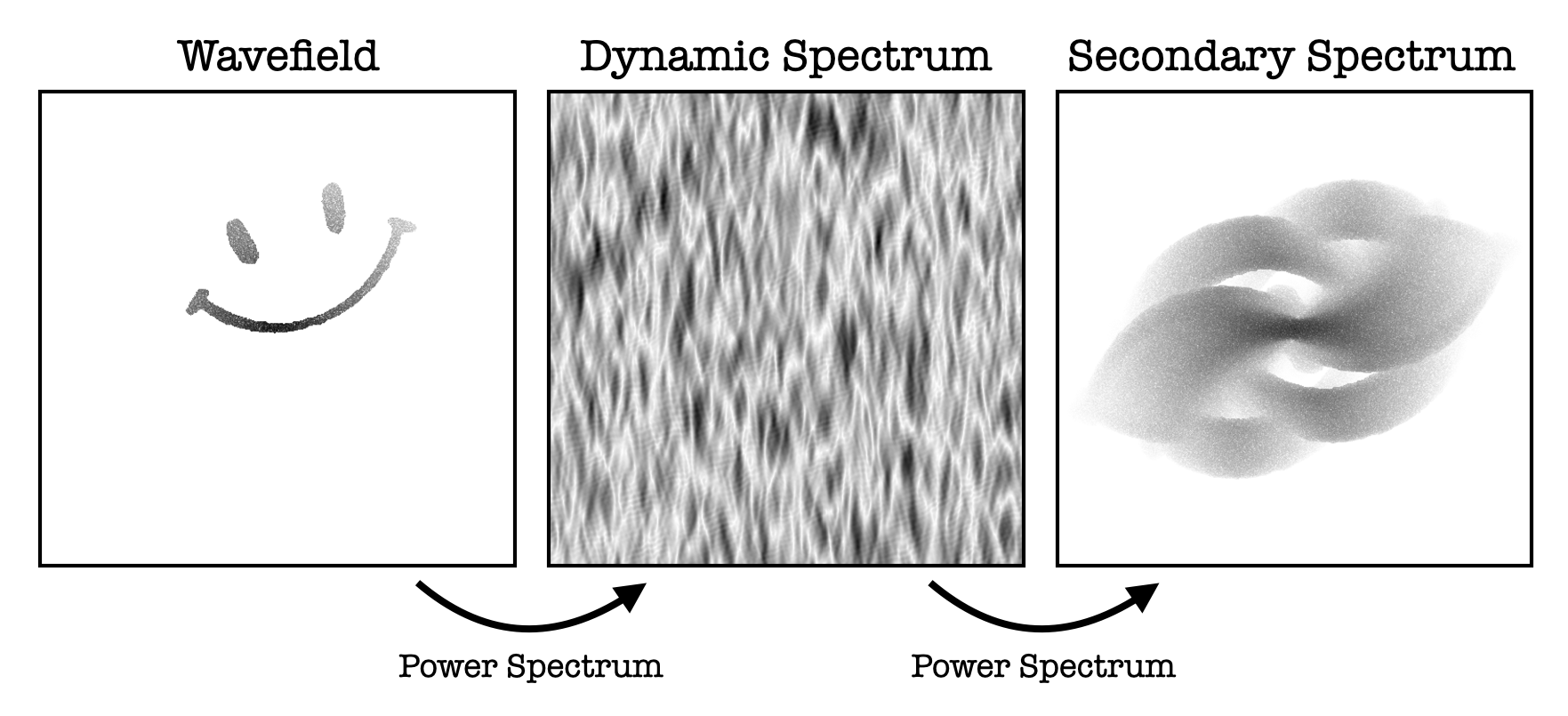}
    \caption{Illustrating the relationships between wavefield, $h(\tau,\omega)= \mathcal{F}\left(H\right)$ (left panel; only the modulus $|h|$ is shown), dynamic spectrum, $D(\nu,t)=|H|^2$ (middle panel), and secondary spectrum $S(\tau,\omega)=| \mathcal{F}\left(D\right)|^2$ (right panel). The secondary spectrum can also be written as $S=|h^* \otimes h|^2$, where $\otimes$ denotes convolution. The dynamic spectrum is the measured quantity in practice, and from it the secondary spectrum is easily evaluated; determining the wavefield requires much more work but the result can offer immediate insights.}
    \label{fig:wavefield_to_secondary}
\end{figure*}

Although secondary spectrum analysis has provided some important insights it is not without its problems. Chief amongst these is the fact that in strong scattering the secondary spectrum can be very complicated and difficult to interpret. The source of this difficulty is easy to spot: the secondary spectrum is actually fourth-order in the wavefield -- i.e. the electric field, $h(\tau,\omega)$, as a function of delay ($\tau$) and doppler-shift ($\omega$) -- being the squared modulus of the convolution of the wavefield with the complex conjugate of itself. The relationships between the wavefield, the dynamic spectrum and the secondary spectrum are illustrated in Fig.~\ref{fig:wavefield_to_secondary}. The main point to observe in this figure is that a simple and highly sparse wavefield yields a complicated and dense distribution of power in the secondary spectrum; the slightly whimsical choice of field structure in this illustration serves to emphasise the clarity of the wavefield as a representation of the signal.

Unfortunately the wavefield is not easy to determine. The problem is that the dynamic spectrum tells us only the dynamic field amplitude, $|H(\nu,t)|$, whereas we also need the phase of $H$ in order to determine the wavefield (via a two-dimensional Fourier transform). By construction, $D(\nu,t)=|H(\nu,t)|^2$, the dynamic spectrum obliterates phases. Another way of thinking about the challenge is to focus on the circularity: the Fourier transform of the dynamic spectrum is $\tilde{D}=h^* \otimes h$, so we need to deconvolve $h^*$ from the observed $\tilde{D}$ in order to arrive at $h$. The process of discovering $h$ from its convolution is prone to error, leading to an imperfect deconvolution, so that our estimate of $h$ is contaminated by features belonging to the conjugate, or ``twin'' image, $h^*$.

Two methods are known that use particular properties of radio pulsar signals to gain access to the wavefield. One is to form a direct estimate of the dynamic impulse response function, $h(\tau,t)$, by extracting sequences of voltages from a baseband recording of a set of giant pulses --- under the assumption that these signals are unresolved impulses at source \citep[][]{2017ApJ...840L..15M}. The wavefield then follows by Fourier-transforming over the sample time, $t$. This approach has the disadvantage that it is restricted in its applicability to the small fraction of pulsars that exhibit giant pulses. The second method exploits the periodic nature of the pulsar signal to construct an intrinsically complex estimator, the cyclic spectrum, which preserves the phase of the signal \citep[][]{2011MNRAS.416.2821D}. The principal disadvantage of cyclic spectroscopy is that it only manifests substantial phases for signals with propagation delays that are large compared to the pulse-width \citep[][]{2013ApJ...779...99W}, whereas we are typically interested in all of the scattered signal components, not just those with large delays. Indeed sometimes the vast majority of the signal is delayed by only a small fraction of the pulse width --- e.g. in the case of nearby, long-period pulsars observed at high radio frequencies. Consequently cyclic spectroscopy, although very helpful, is not a panacea.

In this paper we consider the limiting case of pulsar dynamic spectra, where there is no explicit phase information in the data and the phases must somehow be ``retrieved'' from measurements of $|H|$ alone. Phase retrieval is a procedure that is encountered in many scientific fields and is known to be a difficult problem \citep[see, e.g., the review by ][]{2015ISPM...32...87S}. The general approach is to impose strong additional requirements on the wavefield, beyond simply matching the data, so that a unique solution can be obtained. Bearing in mind that the dynamic spectrum consists of $N$ real numbers, while the dynamic field is described by $N$ complex numbers, it is clear that we need an absolute minimum of $N$ additional constraints in order to make a unique solution possible (number of unknowns equal to the number of constraints). And in practice, with noisy data, it is necessary to have a lot more constraints than unknowns in order to obtain a good solution.

For many years the hybrid input-output (HIO) map of \citet{1982ApOpt..21.2758F} was the principal workhorse for phase retrieval. HIO is an iterative algorithm that utilises alternating projection operators to approach the solution: one projection, the Fourier modulus projection, forces the iterate to match the data, while the other, the support constraint, forces it to lie within a small, predefined region of the solution space. It is now known that HIO is one specific example of a broader category known as Difference Map algorithms \citep[][]{2003JOSAA..20...40E}, in which a variety of constraints can be formulated as projection operators and used in conjunction with the Fourier modulus projection.

In our application we automatically have a support constraint: any scattered signal must have non-negative delay relative to the direct line-of-sight, so $h(\tau,\omega)=0$ $\;\forall\; \tau<0$. This provides us with $N/2$ complex constraints, in addition to the $N$ real data constraints, resulting in exactly as many constraints as unknowns --- barely sufficient to obtain a solution even in the case of zero measurement noise. If it happens that the scattering takes place in only one plane on the sky then our solution space need only be one-dimensional, and in this situation the model wavefield is very tightly constrained by the two-dimensional dataset \citep[][]{2022MNRAS.510.4573B,2022arXiv220413985S}. In the more general case of a two-dimensional scattered image we need to impose additional, strong restrictions on the solution. A tight support constraint cannot be employed because we don't have prior knowledge of where the signal is located, so HIO is not a natural choice of phase retrieval algorithm in this context. But we can instead look for a sparse solution, as was done by \citet{2005MNRAS.362.1279W} and \citet{2008MNRAS.388.1214W} using a method fashioned after the CLEAN algorithm.\footnote{The CLEAN algorithm \citep[][]{1974A&AS...15..417H} is commonly used for deconvolution in the context of interferometric imaging in radio astronomy.} In this paper, rather than CLEAN-ing, we take advantage of recent developments in large-scale, sparse optimisation; specifically, the Fast Iterative Shrinkage Thresholding Algorithm (FISTA) of \citet{10.1137/080716542} \citep[see also][]{10.5555/32048710.5555/3204879}, which we adapt to the context of phase retrieval for pulsar dynamic spectra. Our approach was inspired by the ``Wirtinger Flow'' phase retrieval algorithm of \citet{2014arXiv1407.1065C}.

In FISTA, sparsity is induced by including the $l_1$ norm ($1$-norm) of the solution as one part of a composite objective function that is minimised, the other part being the sum of squared-differences between model and data. The level of sparsity in the solution is controlled by the level of regularisation, i.e. the weight that is given to the $l_1$ norm in the composite objective. Clearly there is a trade-off here: high levels of regularisation help to define a unique solution, but if that solution is too sparse then it cannot provide an accurate model of the data; on the other hand insufficient regularisation may lead to solutions for $h$ that are strongly contaminated by the twin (conjugate) image, $h^*$. To deal with this issue we do not construct our solution via a single optimisation, but rather a sequence of FISTA optimisations using progressively lower levels of $l_1$ regularisation. In this way we seek to navigate to a model of the wavefield that is free of contamination by the twin image, yet is as detailed and accurate as possible --- i.e. limited by the measurement noise in the data. We refer to this approach as ``hierarchical-FISTA'', or H-FISTA for short. We provide the source code of our implementation publicly\footnote{\url{https://github.com/sosl/H-FISTA}}.

This paper is structured as follows. In the next section we set out the details of the hierarchical FISTA approach, as we have designed it, including some explanation of the rationale for the particular choices that were made in this design. The performance of H-FISTA on synthetic test data is shown in \S3, including both noiseless and noisy cases. Although the noiseless case is artificial it is nevertheless of interest in that it represents a limiting condition; perfect reconstruction is shown to be possible in this case. In \S4 we apply H-FISTA to real pulsar data, with all the potential complications that brings (e.g. radio-frequency interference). We demonstrate the results of applying H-FISTA to two observed dynamic spectra: one from the slow pulsar J0837+0610, for which the scattering is practically one-dimensional, and thus the wavefield is highly sparse; the other from the millisecond pulsar J1939+2134, which exhibits a dense wavefield and is therefore less well matched to our method of solution. Discussion and conclusions follow in \S~\ref{discussion} and \ref{conclusions}.

\section{Modelling with hierarchical FISTA}
\label{H_FISTA}

We begin by describing how a sparse model is obtained using $l_1$ regularisation in the FISTA algorithm, before turning to our hierarchical implementation in which the strength of the regularisation is progressively decreased.

\subsection{Obtaining a sparse model with regularisation}
\label{FISTA}

A commonly used method for obtaining sparse models is to employ $l_1$  regularisation when fitting to data --- an approach that is sometimes referred to as the least absolute shrinkage and selection operator or LASSO. This method originated in geophysics \citep{doi:10.1137/0907087} and was independently rediscovered in statistics by \citet{doi:10.1080/00401706.1995.10484371}. It is also widely used in some machine learning algorithms; for example, it is used to avoid overfitting and to perform feature (i.e., parameter) selection \citep[e.g.,][]{2019npjCM...5...83S}.

The method works by including an additional term in the demerit function\footnote{The demerit function is also called cost, loss, or error function}, equal to the $l_1$ norm of the model parameters:
\be
\label{demerit_l1}
F = f\; +\; \lambda\,||h||_1,
\ee
where $h$ represents the model, $\lambda$ ($\ge0$) is the regularisation factor, $||\cdot||_1$ is the $l_1$ norm, $F$ is the total demerit function, and $f$ is a measure of the goodness of fit --- typically the sum of squared differences between model and data. Inclusion of the $l_1$ norm induces sparsity in the model because non-zero components contribute significantly to the demerit even if they are small.

An important consequence of including the $l_1$ term in the demerit function is that the latter is no longer differentiable, which renders many traditional methods of optimisation not usable. Instead, the proximal gradient method must be used, as described in the excellent monograph by  \citet{10.5555/32048710.5555/3204879}. The proximal gradient method relies on the existence of a well-defined and easily calculable proximal operator appropriate to the non-differentiable part of the demerit. In our case, where that part is the $l_1$ norm, the appropriate operator is the soft-thresholding operator, leading to the iterative shrinkage-thresholding algorithm \citep[ISTA,][]{10.5555/32048710.5555/3204879} for minimising $F$.

Each iteration of ISTA yields an updated model, $h_{k+1}$, from that at the previous iteration, $h_k$, by stepping down the gradient of $f$ and then applying the proximal operator to the result:
\be
h_{k+1} = {\rm prox}\left(h_k-\frac{1}{L}\nabla f\left(h_k\right)\right),
\label{ISTA_update}
\ee
where $L$ is the appropriate value of the Lipschitz constant and $\nabla$ is the gradient with respect to $h$. Examining the argument of the ${\rm prox}$ operator in equation~\ref{ISTA_update}  we recognise that the update follows a form similar to that of Newton's method for root-finding, with $L$ playing the role of a curvature of $f$ with respect to $h$. In \S2.3.1 we explain the role of the Lipschitz constant in more detail.

For the demerit given in equation~\ref{demerit_l1}, the ${\rm prox}$ operator in equation~\ref{ISTA_update} is soft-thresholding at threshold $\lambda/L$, defined by:
\be
\label{prox_op_eq}
{\rm prox}\!\left(h\right) \equiv {\rm max}\left(|h|-\frac{\lambda}{L}, 0\right) \frac{h}{|h|}.
\ee
This definition is suitable for our application in which the model, $h$, is complex. We note that using any non-trivial level of regularisation ($\lambda > 0$) leads to values of the model parameters that are biased away from the best fit (minimum $f$) for a given support. On the other hand, when we set $\lambda = 0$, the ${\rm prox}$ operator becomes the identity operator and ISTA can thus be used for unregularised optimisation.

In practice we do not use ISTA, but FISTA --- the fast iterative shrinkage-thresholding algorithm of \citet{10.1137/080716542} \citep[see also ][]{10.5555/32048710.5555/3204879}. FISTA is a first order method, requiring gradient evaluations but not curvatures; it was inspired by the work of \citet{Nesterov1983AMF}. As the name suggests, FISTA typically converges rapidly. On convex problems, i.e. where the goodness of fit is a convex function of the model parameters, the excess demerit (relative to its minimum value) shrinks with iteration number, $k$, at least as rapidly as $1/k^2$. This is much faster than ISTA, which converges as $1/k$. Phase retrieval is known to be a non-convex problem \citep[][]{2002JOSAA..19.1334B}, so in our case these guaranteed convergence rates do not apply.

Rather than the ISTA update given in equation~\ref{ISTA_update}, the FISTA model update is evaluated from a sequence of auxiliary points, $y_k$, that lie close to, but distinct from the model:
\be
\label{FISTA_eq}
h_{k+1} = {\rm prox}\left(y_k-\frac{1}{L}\nabla f\left(y_k\right)\right).
\ee
The sequence of auxiliary points is in turn specified by
\be
\label{point_eq}
y_{k+1} = h_{k+1}+\frac{(-1+t_k)}{t_{k+1}}\left(h_{k+1}-h_k\right),
\ee
in terms of the model, $h$, and the scaling parameter
\be
\label{step_size_eq}
t_{k+1} = \frac{1+\sqrt{1+4\,t_k^2}}{2}.
\ee
To begin the sequence we must choose an initial model, $h_0$, and the other two variables are initialised thus: $y_0=h_0$; and, $t_0=1$.

The fact that FISTA only requires evaluation of the gradient at each step gives it a big speed advantage over methods that require also evaluation of elements of the Hessian (curvature terms); thus FISTA is a good choice for large-scale optimisation problems such as we address in this paper.

Because $f$ is real valued, it cannot be an analytic function of the model wavefield parameters, which are complex, and thus derivatives as they are usually defined on the complex plane do not exist. Instead we use Wirtinger derivatives to form the gradient, as described in appendix \ref{Wirtinger}.

It is important to be aware that FISTA's progress to a solution is in general not monotonic. This behaviour comes about because the iteration, viewed as a dynamical evolution, includes a ``momentum-like'' term -- the second term on the right-hand-side of equation~\ref{point_eq} -- which causes the trajectory to overshoot when it passes close to the minimum \citep[][]{JMLR:v17:15-084}. If a monotonic sequence is desired it can be achieved by restarting FISTA, with $h_0$ taken as the model corresponding to the minimum demerit achieved on the current trajectory \citep[][]{10.5555/32048710.5555/3204879}. For our application it is unimportant whether or not the evolution is monotonic and therefore we have not employed restarting.

To summarise, the FISTA algorithm is:
\medskip

\begin{tabular}{ll}
{\bf Input:} & $L$ - the Lipschitz constant \\
                 & $\lambda$ - regularisation parameter \\
                 & $h_0$ - initial model of the wavefield\\
{\bf Step 0:} & $t_0 = 1$   \\
              & $y_0 = h_0$ \\
              
{\bf Step k+1:} & $h_{k+1} = {\rm prox}\left( y_k - \frac{1}{L} \nabla f (y_k)\right)$  \\
  & $t_{k+1} = \frac{1+\sqrt{1+4\,t_k^2}}{2}$ \\
  & $y_{k+1} = h_{k+1} + \frac{(-1+t_k)}{t_{k+1}}\left(h_{k+1} - h_k\right)$ \\
\end{tabular}

\subsection{The demerit function}
We seek a model wavefield, $h(\tau,\omega)$, that fits the data, i.e. the observed dynamic spectrum $D(\nu,t)$. The model dynamic spectrum is simply
\be
\label{predicted_ds}
Z(\nu,t) = H(\nu,t)\,H^*(\nu,t),
\ee
in terms of the dynamic field,\footnote{This quantity is called the filter in \citet{2013ApJ...779...99W}.} $H$, where 
\be
H = \mathcal{F}^{-1}\left(h\right),
\ee
and $\mathcal{F}^{-1}\left(\cdot\right)$ denotes the inverse Fourier transform. We thus form the residual, $R=Z-D$, between model and data, and the goodness of fit is gauged by the sum of squared residuals:
\be
f(h)\equiv \frac{1}{2} \sum_{\nu,t} R^2.
\ee
And finally we have the real-valued demerit function which depends on the complex-valued wavefield $h$:
\be
\label{full_demerit_scalar_lambda}
F(h) = f(h)+\lambda\,||h||_1
\ee

In practice it is convenient to implement different levels of regularisation for different regions of the modelling space, as follows. The solution we seek lies in the positive delay half-space, so we exclude negative delay components from the model by imposing regularisation with $\lambda = \infty$ on the region $\tau<0$. In the positive delay half-space, by contrast, we have a finite value for $\lambda$. A small fraction of the positive delay components have already been established (by previous FISTA optimisations) as a necessary part of the model, so they are optimised without regularisation ($\lambda = 0$), in order to avoid bias --- these are referred to as ``approved'' components in this manuscript. Consequently the demerit function that we use in practice has the form
\be
\label{full_demerit_lambda}
F(h) = f(h)+ ||\Lambda\odot h||_1,
\ee
where $\Lambda$ is a matrix of regularisation factors, and $\odot$ indicates the Hadamard product (i.e., element-wise multiplication); and each element of $\Lambda$ takes one of three values --- $0$, $\lambda$, or $\infty$. Similarly, we replace the scalar regularisation factor $\lambda$ in the ${\rm prox}$ operator (\ref{prox_op_eq}) with the matrix $\Lambda$.\footnote{After developing this approach we became aware that non-uniform $l_1$ regularisation has previously been proposed in bioinformatics by \citet{10.1093/bioinformatics/btaa776}, who refer to the method as ``differential shrinkage'', and their whole algorithm as ``xtune LASSO''.}

\subsection{Backtracking to the Lipschitz constant}\label{backtracking}
The guaranteed convergence properties of FISTA on convex problems, mentioned in the previous section, depend on knowing the appropriate value of the Lipschitz constant for $\nabla f$. It is determined by the requirement
\be
||\nabla f\left(y\right) - \nabla f\left(x\right)||\; \le \; L\, || y - x ||,
\ee
for all possible choices of $x,y$ within the region of interest. For some particular examples of functions, $f$, it is possible to evaluate $L$ analytically, but we have not been able to establish an analytic result in our case. Instead we proceed by making an initial estimate of $L$, as described in Appendix \ref{Lipschitz_calculation}, and then at each iteration of FISTA we check for consistency; this procedure is known as ``backtracking'' \noindent\citep[][]{10.1137/080716542,10.5555/32048710.5555/3204879}.

If we make an estimate of $L$ that is too large then FISTA will converge more slowly than it would with an accurate value of $L$. On the other hand, if the current estimate of $L$ is too small then FISTA may not converge at all --- which is a much more serious problem. Consequently backtracking is simply designed to ensure that our estimate of $L$ is not too small, as gauged by the requirement that $f$ is bounded locally by a quadratic form \citep[][]{10.1137/080716542,10.5555/2670022,10.5555/32048710.5555/3204879}. For our case, where the domain is complex, the appropriate bound is
\be
\label{backtracking_eq}
f(x) \le f(y)+2\Re\left(\langle\left(x-y\right)^*, \nabla f(y)\rangle\right) + \frac{L}{2} ||x-y||^2,
\ee
where $\Re\left(\cdot\right)$ indicates the real part, and $\langle\cdot\,, \cdot\rangle$ is the dot product. We check whether this inequality is satisfied at each step of the FISTA iteration; if not we increase $L$ by a factor $\eta>1$ (we have adopted $\eta=1.15$) and reevaluate, repeating this process until the inequality (\ref{backtracking_eq}) is satisfied. 

\subsubsection{Initial wavefield and support constraint}
\label{init_h_support_constraint}
A number of strategies exist for choosing the initial model in iterative approaches to phase retrieval, as described in \citet{2020arXiv200405788F}. These methods include spectral initialisation, null initialisation, pre-processing and random initialisation. In our case we have very little a priori information on the underlying wavefield so we use a simple initial guess: a single component model at the origin ($\tau=0=\omega$). This choice corresponds to a constant predicted dynamic spectrum $Z$.

We initialise the value of the only non-zero component in the initial wavefield $h_0$ such that $Z = \bar{D}$ where $\bar{\cdot}$ indicates the average. That is, our initial wavefield model reproduces the average value of the observed dynamic spectrum without any further structure. As we are confident that this component should be present in the wavefield -- i.e. it is automatically an approved component -- we set the corresponding element in the $\Lambda$ matrix to zero, thus ensuring that we recover an unbiased estimate of this component during optimisation.

As usual with phase retrieval problems, there is an ambiguity in our model as can be seen from equation~\ref{predicted_ds}: if we rotated the phase of the wavefield by some constant value, and thus also rotated the phase of the dynamic field, the dynamic spectrum model, $Z$, would not change. In other words: multiplying $H$ by $e^{i\phi}$ does not affect $Z$. Because of this, the wavefield $h$ can rotate in overall phase during the fitting, and to prevent that from happening we force the phase at the origin of the wavefield to be zero at all times.
    
In contrast to the secondary spectrum, the wavefield $h$ cannot contain any signals at negative delays as they are not causal. This provides a weak support constraint for our phase retrieval problem. To implement this constraint in our algorithm, we set the negative delay regions of the regularisation matrix, $\Lambda$, to infinity, thus ensuring that the soft-thresholding operator will always set the corresponding wavefield coefficients to zero.

In addition to the familiar ambiguity in the overall phase of the wavefield, the predicted dynamic spectrum (unlike the cyclic spectrum) is insensitive to the location of the origin of the wavefield coordinates.\footnote{To see why: if we multiply our model $H$ by the phase factor $\exp[i(\tau_0\nu+\omega_0 t)]$, the model $Z$ is unchanged.}
This degeneracy, in combination with a causal support constraint, can introduce problems, as follows. Our initial, plane wave model naturally represents the strongest individual plane wave component in the real wavefield, no matter where it occurs. However, that strongest component might actually be located at delay $\tau_0>0$, so that any bona fide signal components in the delay range $0\le \tau < \tau_0$ will be eliminated from our model if we use a strictly causal support constraint. To mitigate this problem we include a region with small negative delays within our support; in our software this region has a default extent of four pixels (but can be adjusted by the user).

\subsubsection{Choice of regularisation level}
\label{initial_lambda}
The character of the solution returned by FISTA is strongly dependent on the numerical value of the regularisation parameter, $\lambda$: if $\lambda$ is set to a sufficiently large value then the optimum wavefield is $h=0$, whereas if we use $\lambda=0$ then there is no regularisation and the result will be a fully dense wavefield that is strongly contaminated by the twin image. Neither extreme is useful. Instead we need to choose a value of $\lambda$ that yields a highly sparse but non-trivial model; we will call this value $\lambda_{\rm init}$. 

Based on testing with both real and simulated dynamic spectra (typically with overall size $N_\nu \times N_t \sim 10^6$) we have found that our hierarchical approach, H-FISTA (see \S2.4), typically performs best if the first FISTA iteration returns a wavefield model having $N_0$ non-zero components with $10 \la\, N_0\, \la 100$. And we can relate $N_0$ to the initial gradient and regularisation level as follows. Excepting the origin, our initial wavefield model is null, so the proximal operator in equation~\ref{ISTA_update}  acts on the quantity $-\nabla f(h_0)/L$. The proximal operator (equation~\ref{prox_op_eq}) yields a null result unless $|h|>\lambda /L$, so the number of non-zero components in the wavefield model after the first FISTA iteration will be the number of components for which
\be
\label{lambda_init}
 |\nabla f(h_0)| > \lambda_{\rm init}.
\ee
Thus once $N_0$ is selected we can choose a suitable value of $\lambda_{\rm init}$: we set $\lambda_{\rm init}$ equal to the $N_0$-th largest value of the modulus of the intial gradient. Unless otherwise specified, in this paper we have employed $N_0 = 60$.

We note that while we can control the exact number of new components present in the wavefield after the first FISTA iteration, we cannot predict how many non-zero components will be present in the model after subsequent FISTA iterations. Empirically we have found that the number of non-zero components in the optimised wavefield model (for regularisation $\lambda=\lambda_{\rm init}$) is typically within a factor $\sim2$ of $N_0$.

\subsubsection{Number of FISTA iterations}
\label{FISTA_niter}
In our approach the FISTA algorithm is used multiple times, with different levels of regularisation, as described in \S2.4. Excepting the first and last such levels, which have special requirements (see \S2.4), the aim at each stage is simply to improve the wavefield model, and for this goal it suffices to use a fixed number of iterations, $N_{\rm iter}$, in each FISTA optimisation. After some experimentation we chose $N_{\rm iter}=80$, as this value seems to work well for both the synthetic and real data we have tested our algorithm on.

\subsubsection{Debiasing and hard thresholding}
\label{debiasing}

After FISTA has gone through $N_{\rm iter}$ iterations, we stop the optimisation. The model wavefield at this stage is affected by a number of issues: a) the total number of non-zero components may be too small to describe the data well b) some twin image components may be present in the model, albeit hopefully only at low levels, and c) the ``new'' components of the image -- i.e. those for which the corresponding value in the regularisation matrix $\Lambda$ was above zero and finite -- are biased by the soft-thresholding operation. We defer addressing the first of these issues, and consider now the other two.

Dealing with bias in the wavefield model is straightforward: we set the elements of $\Lambda$ to zero where the wavefield is non-zero (and $\infty$ elsewhere), and run FISTA again. In this case our ${\rm prox}$ operator is the identity (or else null), and we are performing a simple (unregularised) least-squares optimisation with a fixed support. After this debiasing step we expect that all legitimate components of the wavefield model ought to exceed the threshold value $\lambda/L$ used in the ${\rm prox}$ operator (3) for the original, regularised FISTA optimisation. We can therefore reasonably exclude any weaker components from the model on the grounds that they are less reliable than the stronger ones and, in particular, they may well be components of the twin image. We therefore apply a hard thresholding operation on the debiased wavefield:
\be
\label{hard_threshold_eq}
h \rightarrow \; T_{\rm H}(h)\,=\, h\,\odot\,\Theta\left(|h|-\epsilon\frac{\lambda}{L}\right),
\ee
where $\Theta$ is the Heaviside step function. Our software includes a scaling parameter, $\epsilon$, for this hard thresholding step that can be adjusted by the user, with $\epsilon=1$ by default.

If any components are removed by the hard thresholding operation, we again perform an unregularised optimisation followed by hard thresholding, and repeat. In our testing, most of the time only one such loop is executed. The remaining non-zero components at this stage are referred to as ``approved'' components, and to avoid introducing unnecessary bias into our wavefield model we set the corresponding elements of $\Lambda$ to zero in all subsequent FISTA optimisations --- as per the comments following equation~\ref{full_demerit_lambda}.

\subsection{Hierarchically extending the wavefield model}
\label{HFISTA_details}
The algorithm described above is designed to yield a model wavefield that is free of contamination by the twin image, and has unbiased component values, but is likely to be too sparse to describe the data accurately. We cannot improve the accuracy of the model by simply increasing $N_0$ (or, equivalently, decreasing the initial regularisation $\lambda_{\rm init}$) without potentially increasing the level of contamination by the twin image. Instead we take our optimised sparse wavefield model as the starting point for a new FISTA optimisation, as already described but with a lower level of regularisation, $\lambda<\lambda_{\rm init}$. And this whole process of optimisation, debiasing, hard-thresholding, and further reducing the regularisation factor is repeated in a loop. We use a geometric sequence of $\lambda$ such that the k-th value of the regularisation factor is:
\be
\label{lambda_sequence}
\lambda_{\rm k} = \frac{\lambda_{\rm init}}{\eta_\lambda^{k}}
\ee
with scaling factor $\eta_\lambda=1.15$. In this way we gradually build a detailed description of the wavefield, while nevertheless imposing significant regularisation on all of the new components as they are added to the model, no matter what their strength.

\subsubsection{Stopping criterion}
\label{FISTA_stopping}
The final piece we need to complete the H-FISTA algorithm is a stopping criterion for looping through the $\lambda$ values. We use two stopping criteria and interrupt H-FISTA when either of them is triggered.

The first criterion, which we refer to as the sparsity criterion, is simple: we stop iterating through $\lambda$ values if the number of non-zero components present in the wavefield exceeds a certain maximum value. This criterion is used to ensure that the wavefield model is indeed sparse, so that the optimisation process is well constrained. In our software the default threshold level corresponds to a wavefield with only $3$ per cent of components being non-zero, but the threshold can be adjusted by the user --- e.g. in order to obtain solutions for denser wavefields.

The second criterion, which we refer to as the spatial criterion, is based on the idea that noise is uniformly distributed over the domain of the solution, whereas the signal is not --- signal components are generally expected to be stronger near the origin, for example. Thus if our $\lambda$ parameter is sufficiently small that we are adding predominantly noise components we expect those new components to be added fairly uniformly across the doppler-shift -- delay space. In our test cases we found that comparing the doppler-shift distribution of new wavefield components with a uniform distribution worked best. Specifically, we perform a Kolmogorov-Smirnov (KS) test if there are at least $100$ new non-zero wavefield components, to determine how closely their doppler-shifts conform to a uniform distribution across the accessible doppler space. If the KS statistic is above a threshold we decide that in the current loop of H-FISTA (i.e., current $\lambda$ value) we added primarily noise, discard the wavefield and use the wavefield obtained with the previous $\lambda$ value as the sparse solution.

We note that the threshold statistic should be quite small, much smaller than what would be acceptable to decide the distribution of the doppler-shift values is indeed uniform. This is because there will often still be a number of components being added that are above the noise and part of the wavefield, and those components are likely to not be uniformly distributed. In other words, we are not trying to capture the moment that the distribution achieves uniformity, but rather when it starts veering towards it. In practice we found that a KS probability value of $10^{-10}$ worked well for our test cases.

With these two criteria defined we run H-FISTA, with progressively decreasing $\lambda$, until either the spatial or the sparsity stopping criterion is met. Our software implementation allows the user to adjust the threshold levels for either of these criteria, or to disable either of them.

\subsubsection{H-FISTA algorithm summary}
Here we summarise the whole H-FISTA algorithm we use to obtain a sparse solution for the wavefield.

\begin{tabular}{ll}
{\bf Input:} & $\lambda_{\rm init}$ - initial regularisation parameter \\
                 & or instead \\
                 & $N_0$ - model components after the very first FISTA step \\
                 & $\epsilon$ - scale factor for hard thresholding operator $T_{\rm H}$ \\
                 & $N_{\rm iter}$ - number of iterations in each FISTA optimisation\\
                 & $\eta_\lambda$ - $\lambda$ scaling factor \\
{\bf Step k:} & \\
  & 1. Set $\lambda_k = \lambda_{k-1}/\eta_\lambda$ \\
  & 2. Set $\Lambda=0$ where $h\ne 0$, $\lambda_k$ elsewhere \\
  & 3. Run FISTA \\
  & 4. Set $\Lambda=0$ where $h\ne 0$, $\infty$ elsewhere \\
  & 5. Run FISTA \\
  & 6. Perform hard thresholding with threshold $\epsilon\lambda/L$ \\
  & 7. If any components zeroed, go back to step 4 \\
  & 8. Check stopping criteria: exit or return to step 1 \\
\end{tabular}

\subsection{Obtaining a dense wavefield}
\label{densify}
By design, H-FISTA aims to capture the vast majority of significant signals in a sparse model, as described above. However, there is no clear boundary between ``signal'' and ``noise'' and users may be interested in wavefield components that are weaker than those identified by H-FISTA. In particular we note that components that are clustered around a certain location in the solution space $(\tau,\omega)$ might be individually insignificant but collectively significant. Consequently we expect that most users will prefer a fully dense wavefield model over the sparse solution that is provided by H-FISTA, and we have therefore provided a mechanism for obtaining a dense model.

Our approach assumes that the sparse solution provided by H-FISTA is close to the true minimum of the demerit. Thus it is likely that the model is constrained well enough that a simple unconstrained FISTA can be used to fit the wavefield to the data. That is, we use FISTA with a fixed number of iterations $N_{\rm dense}=1000$ and with $\lambda=0$ to obtain a dense solution. This number of iterations is likely to be sufficient as we start close to the minimum demerit, and can be adjusted as needed. The result is a fully dense model of the wavefield, with non-zero values at all locations $(\tau,\omega)$ --- even at negative values of the delay.

\subsubsection{Experimental application of the difference map to densification}
As mentioned in the introduction, the difference map algorithm is commonly used for phase retrieval, but is not well suited to our context because we have only very weak a priori support constraints (i.e. causality). We did, however, try using the difference map as a method of arriving at a dense wavefield, starting from the sparse solution obtained with H-FISTA. At this stage a support constraint is inappropriate because we want to obtain a non-zero estimate for every point in the delay-doppler plane, but a strong constraint of some kind is necessary in order to obtain a meaningful solution. By design the sparse model arrived at with H-FISTA should contain all of the statistically significant components of the wavefield, with the remaining values being predominantly noise. In this circumstance we can implement a histogram projection,\footnote{This operator is a projection that forces components to follow a particular statistical distribution} as described by \citet{2003JOSAA..20...40E}, because we know the probability distribution function for the noise.

In practice we found that this method did not perform as well as the simpler approach described in the previous section: we found that employing a noise-histogram projection suppressed the remaining, weak signals in the wavefield --- clearly an undesirable result. For that reason we ultimately decided not to use the difference map for densification.

\section{Application to synthetic data}
\label{synth_dynspec}
We now demonstrate the usefulness and limitations of our method using synthetic dynamic spectra. We start with an example of a noise-free dynamic spectrum to demonstrate a perfect recovery of a sparse wavefield using H-FISTA. We show the second case of a noise-free data and use it to demonstrate some of the limitations of our method. For a more realistic demonstration, we provide a third example in which we add random noise to the wavefield recovered in the noise-free case and use the resulting dynamic spectrum as input data for H-FISTA.

\subsection{Noise-free sparse wavefield}
\label{noise_free}

\subsubsection{Example of successful recovery}
\label{noise_free_success}

\begin{figure}
    \centering
    \includegraphics[width=\columnwidth]{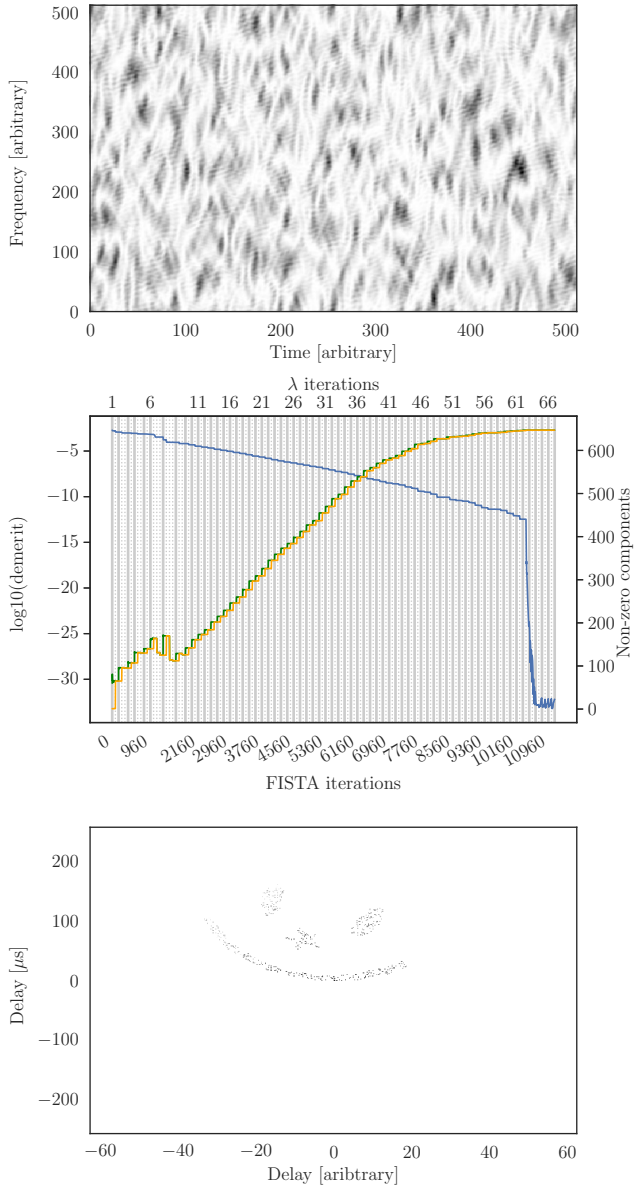}
    \caption{Successful application of H-FISTA to simulated, noise-free data. Top panel: the input dynamic spectrum, shown with inverted grey-scale intensity. Middle panel: diagnostics of the H-FISTA loops as a function of the total number of FISTA iterations and $lambda$ iterations. The blue line shows the logarithm of the demerit, the green line shows the number of non-zero components in the wavefield model, and the orange line shows the number of ``approved'' components. Thicker, continuous grey lines indicate a change of $\lambda$ value, while the thinner, dotted grey lines mark the boundaries between sequences of unregularised FISTA iterations. Bottom panel: the recovered wavefield, shown as $\log_{10}\left(|h|^2\right)$. Essentially perfect recovery of the wavefield is achieved in this case, with the number of components reaching a plateau and the demerit limited by machine precision.}
    \label{fig:dynspec_smiley_125_noisefree}
\end{figure}

As a limiting case, we start by generating a sparse wavefield with $647$ non-zero components and no noise, i.e., all the other components are exactly zero. The location of the non-zero components was selected randomly within a mask consisting of four separate areas at positive delays. Of all components within the mask, one in eight was assigned a non-zero complex value, with both phase and modulus drawn randomly. The corresponding dynamic spectrum is shown in the top panel of Fig.~\ref{fig:dynspec_smiley_125_noisefree}.

Using this dynamic spectrum as input, we run the H-FISTA algorithm as described in \S~\ref{H_FISTA}. We note that in the noise-free case the stopping criterion will not work if we successfully recover the sparse wavefield as there will never be any components with uniformly distributed doppler-shifts in the solution. The sparsity-based criterion will also not interrupt the loop as the total number of components in the wavefield is below the default threshold. Instead, we used 80 iterations per FISTA run and adjusted the regularisation parameter $\lambda$ by $\eta_\lambda=1.15$ for every step of the outermost loop of the algorithm until the number of non-zero components in the wavefield stabilises. To ensure the stability of the algorithm, we ran a number of additional iterations through regularisation level $\lambda$ to check if the solution remains stable.

In this case, we successfully recover the input wavefield and show it in the bottom panel of Fig.~\ref{fig:dynspec_smiley_125_noisefree}. The middle panel shows an H-FISTA diagnostic plot which shows several quantities.

The blue line is the base 10 logarithm of the demerit function. The green line indicates the total number of non-zero components of the wavefield. The orange line is the number of approved components, i.e., components for which the regularisation parameter is $\lambda=0$. The continuous vertical grey lines indicate a change of $\lambda$ value while the thinner dotted grey lines indicate an end of a FISTA run and a hard thresholding episode without adjusting $\lambda$. 

All these quantities are shown as a function of the global FISTA iteration count. We note only some of the labels are provided to avoid overlap.

A few features are notable in this diagnostic plot. First of all, for a fixed $\lambda$ value, we typically see a significant increase in the number of all components which is expected as the regularisation allows more components after a new value of $\lambda$ is selected.

Secondly, most of the H-FISTA loops consist of only two FISTAs, one with a specific value of $\lambda$ and one debiasing run. In other words, most of the time, one application of hard-thresholding is sufficient and it is common to see no components being removed by this process. We also see that the number of components often drops during FISTA iterations as the soft thresholding can reduce them to zero. The overall trend of the number of components is not monotonic.

The third insight is that there are H-FISTA loops with significant hard-thresholding episodes. These occur relatively early on, around a 1000 FISTA steps into the phase retrieval. These are crucial for removing the twin image from the wavefield estimate and such episodes are often seen for the successful retrievals.

One feature visible in this diagnostic plot is specific to the noise-free case. Namely, the number of components stabilises at a fixed value during the iteration number $62$ of H-FISTA and the demerit plummets rapidly once all the components are correctly identified and debiased. Such behaviour would not be seen in cases with noise. The solution remains stable after finding all the components of the input wavefield.

Finally, we note that due to the large dynamic range of the plot the demerit appears to monotonically decrease through the whole procedure. As we noted earlier, FISTA does not guarantee that the demerit behaves this way. There are oscillations of the demerit present in this figure, however, they are too small to see with such a large dynamic range.

\subsubsection{Failed phase retrieval for noise-free case}
\label{noise_free_fail}

Here we present an attempt at recovery of the wavefield from a noise-free dynamic spectrum generated similarly as it was the case above The only difference is that instead of one in eight, we now allow twice as many components within the pre-selected regions to be non-zero. That is, within the four regions of the wavefield seen in the bottom panel of Fig.~\ref{fig:dynspec_smiley_125_noisefree}, one in four components are non-zero, resulting in a wavefield that is less sparse locally.

\begin{figure}
    \centering
    \includegraphics[width=\columnwidth]{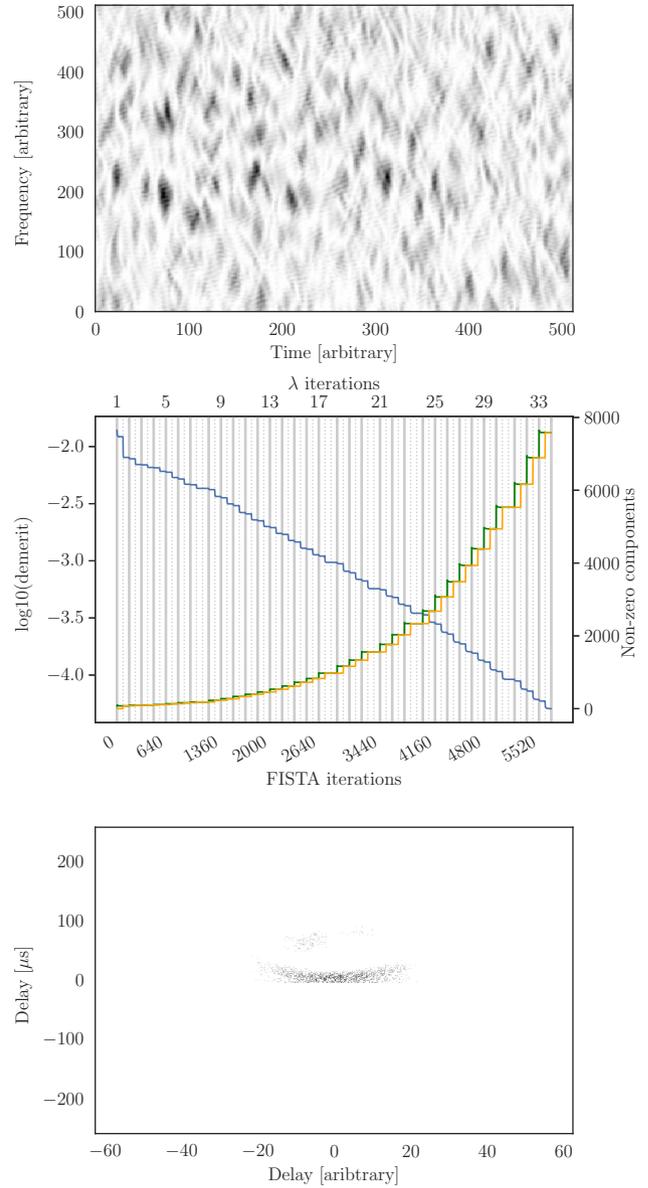}
    \caption{As figure~\ref{fig:dynspec_smiley_125_noisefree}, but with twice the number of non-zero components in the input wavefield (see text, \S~\ref{noise_free_fail}). In contrast to figure 2 we can see that: (i) the number of components in the model does not reach a plateau, but continues to rise until the process is terminated; and, (ii) the best demerit achieved is approximately eight orders of magnitude larger than the limit set by machine precision. The resulting model is locally dense, and is a poor rendition of the input wavefield. The lower panel shows the model obtained after $20$ H-FISTA iterations rather than for the last iteration.}
    \label{fig:dynspec_smiley_25_noisefree}
\end{figure}

Using the dynamic spectrum generated from this wavefield (shown in the top panel of Fig.~\ref{fig:dynspec_smiley_25_noisefree}) we run H-FISTA using the same configuration as before. In this case, the algorithm never converges and the number of components would keep increasingly indefinitely if we disabled the stopping criteria. Eventually, the whole available parameter
space would be occupied by non-zero components. Our diagnostic plot shown in the middle panel of Fig.~\ref{fig:dynspec_smiley_25_noisefree} shows the diagnostic extending until the end of 33rd H-FISTA iteration. The sparsity stopping criterion is triggered for the iteration number $34$ and we discard that iteration. The other crucial difference between this diagnostic and that for the previous case is that the hard thresholding episodes are not as significant with only a small number of components set to zero. As a result, the twin image is never fully removed and the wavefield recovery does not succeed.

In the bottom panel Fig.~\ref{fig:dynspec_smiley_25_noisefree}, we present the wavefield during one of the H-FISTA steps to illustrate this failure mode. Specifically, this solution was obtained after $20$ iterations. We chose this particular wavefield as at this step the wavefield has $1558$ non-zero components, which is close to the true number of components in the wavefield from which we generated the input dynamic spectrum. Three of the four regions have non-zero components in the wavefield solution at this stage. However, there are too many components present. This problem is most evident in the crescent-shaped region. The region now extends beyond the boundaries visible in the wavefield recovered in the successful case (bottom panel of Fig.~\ref{fig:dynspec_smiley_125_noisefree}). Furthermore, there are components present even at the most negative delay we allow. As $\lambda$ gets reduced further and further, the wavefield becomes more and more densely populated and eventually all the allowed components have non-zero values (not shown in the plots).

\subsection{Sparse wavefield with noise}
\label{synth_with_noise}

\begin{figure}
    \centering
    \includegraphics[width=\columnwidth]{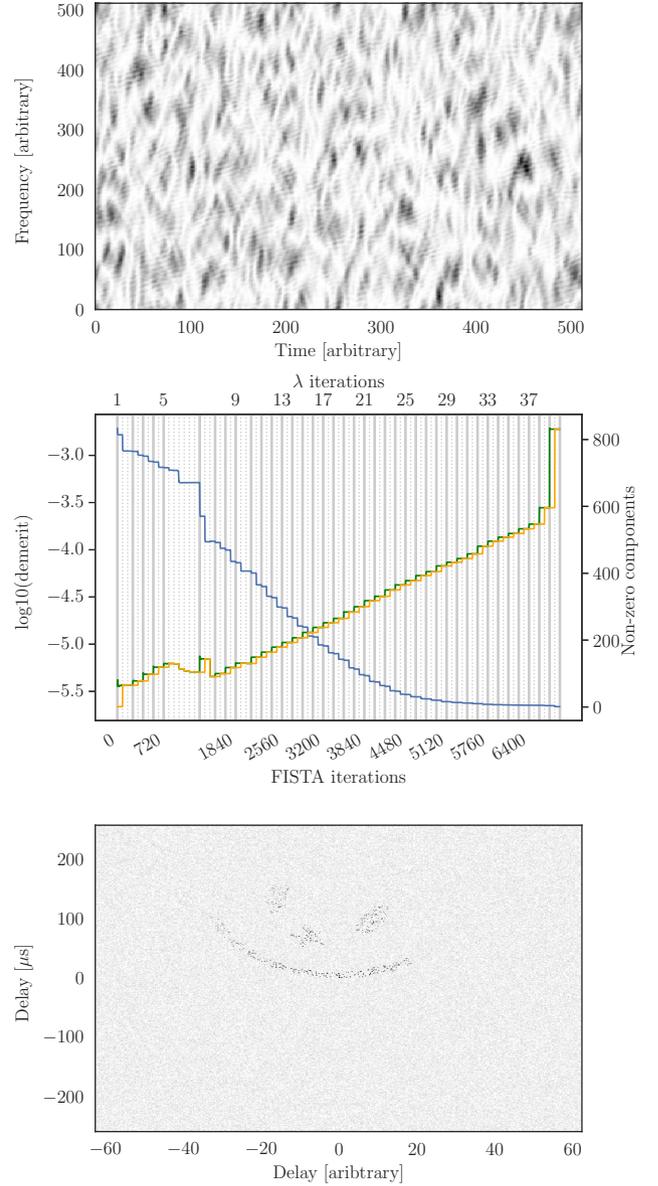}
    \caption{As figure \ref{fig:dynspec_smiley_125_noisefree}, but with noise added to the input wavefield (with the transfer function chosen for contrast on significant components of the wavefield). In this case on the middle panel we extend the diagnostic plot to include the iteration on which the stopping criterion was triggered for demonstration purposes. We note that the diagnostic plot includes H-FISTA steps beyond the stopping criterion for demonstration purposes.}
    \label{fig:dynspec_smiley_125_noisy}
\end{figure}

We now move on to a case of a synthetic dynamic spectrum based on a sparse wavefield to which noise was added. That is, we have a dense wavefield, but the actual signal is sparse. To simulate this, we take the sparse wavefield used in \S~\ref{noise_free_success} and add random normally distributed noise to the real and imaginary parts of every wavefield component. The added noise has zero mean and variance is chosen such that noise overlaps with the low end of the signal components and about $90$ per cent of the components of the input noise-free wavefield are still above the highest modulus of the noise realisation. The resulting dynamic spectrum is visually similar to what is shown in Fig.~\ref{fig:dynspec_smiley_125_noisefree}, and is shown in the top panel of Fig.~\ref{fig:dynspec_smiley_125_noisy}.

We run H-FISTA now with default settings on this input dynamic spectrum. The only difference in the method compared to what we did in \S~\ref{noise_free_success} is that we now include the stopping criteria as we expect them to work properly in a case with noise. Furthermore, we will calculate a dense solution which we did not do in the noise-free case.

If we add too much noise, more components within the regions with a signal are at a level comparable to the signal, thus turning this recovery problem into a similar problem to what we have seen fail in \S~\ref{noise_free_fail}. In such cases, the resulting wavefield looks similar to what we saw in the bottom panel of Fig.~\ref{fig:dynspec_smiley_25_noisefree}.

    However, with the amount of noise specified above, the phase retrieval succeeds. The diagnostic plot is shown in the middle panel of Fig.~\ref{fig:dynspec_smiley_125_noisy}. Compared to the noise-free case for which the phase retrieval failed, there are more hard-thresholding episodes. The demerit also drops sharply during one of the iterations which was not seen in the middle panel of Fig.~\ref{fig:dynspec_smiley_25_noisefree}. Compared to the noise-free version of the same underlying wavefield, there is no sharp drop of the demerit towards extremely low values, as the combination of sparsity and the presence of noise prevents a perfect fit to data.

The diagnostic figure includes the 39th H-FISTA iteration which triggered the spatial stopping criterion. We remind the readers that this iteration is normally discarded and the sparse solution is the wavefield estimated with the second last regularisation factor $\lambda$. We include it on the diagnostic plot in this case to demonstrate the sharp increase in the number of components added which is often the case when H-FISTA starts adding predominantly noise components.

H-FISTA iterations are interrupted by the spatial stopping criterion. The sparse wavefield consists of $596$ non-zero components. While the number of non-zero components is very close to the number of components in the noise-free version of the wavefield, we note that some of the components retrieved are noise components and some of the signal components are missing. We use this sparse solution to derive the dense wavefield as described in \S~\ref{densify}. The resulting wavefield is shown in the bottom panel of Fig.~\ref{fig:dynspec_smiley_125_noisy}. 

The retrieved wavefield is consistent with the noise-free input wavefield (shown in the bottom panel of Fig.~\ref{fig:dynspec_smiley_125_noisefree}).

\section{Application to observed dynamic spectra}
\label{observed_dynspec}

In \S~\ref{synth_dynspec} we demonstrated that H-FISTA can successfully retrieve the phases from some idealised synthetic dynamic spectra, both in the noise-free case and for data with low levels of gaussian noise. In this section we apply H-FISTA to analyse observed dynamic spectra. Three major differences arise between the observed and simulated the dynamic spectra: we generally do not know the answer upfront, the data may be affected by radio frequency interference (RFI), and the data may be affected by intrinsic variability with deleterious effects on the phase retrieval in our implementation.

To provide a degree of verification, we use dynamic spectra for which the wavefields are available in the literature. Before we move on to applying our method to observed dynamic spectra, we address the latter two issues.

\subsection{Handling data imperfections}
\subsubsection{Dealing with radio frequency interference}
\label{RFI}

Radio observations of pulsars are frequently affected by RFI. Here we deal mostly with narrowband RFI and our method follows the standard median filtering used in the pulsar community \citep[see e.g.,][]{2012AR&T....9..237V}.

To identify RFI in the dynamic spectrum, we calculate a running median in a window of 21 samples. We then estimate the median absolute deviation (MAD) of the data and mark anything more than five times the MAD as RFI. This process is repeated at the end of every loop of H-FISTA (that is, once per value of $\lambda$) and the RFI mask is updated.

We use the RFI mask to exclude affected parts of the dynamic spectrum from the calculation of the residuals, estimation of the Lipschitz constant, and estimation of the gradient. We note that our RFI mask will often flag channels near the edges of the observed band, where the bandpass response rolls off, effectively trimming the dataset (if not already done as part of the data pre-processing).

Even if the RFI mask generated covers more of the dynamic spectrum than expected, we recommend proceeding as is and not tuning the zapping algorithm as H-FISTA appears reasonably robust to losing a small fraction (i.e., a few per cent) of the input data.

If the data contains gaps in time domain, e.g., due to a stop in observation, the same method can be used to generate a mask to remove these gaps by applying it to frequency averaged data instead.

Finally, we note that the users can provide a pre-existing mask of RFI as input to the algorithm. Any further RFI mitigation in H-FISTA can also be disabled if desired.

\subsubsection{Dealing with intrinsic variability}
\label{intrinsic_variability}

Another potential issue with data which can create difficulties with phase retrieval using H-FISTA as formulated above, is variability of flux of the pulsar, either intrinsic or external (e.g., due to instrumental causes such as variable gain). This can show up as a high-frequency changes of the intensity, even every time sample of the dynamic spectrum.

In our case, we first estimate if this is likely to be an issue for the input dynamic spectrum by visual inspection of the spectrum itself, as well as of the Fourier transform of the frequency average of the dynamic spectrum. In the dynamic spectrum space, the need for filtering of the variability shows up as a striation.

If the inspection reveals the presence of striation, we calculate the frequency average of data and normalise the full-resolution data by that average. This method has been used by, e.g., \citet{2019MNRAS.485.4389R} and \cite{2022A&A...663A.116W}. With the striation largely removed from the dynamic spectrum, we do not include this variability in our model. 

Finally, we note that the data can be pre-processed in any other way before using it as input for H-FISTA. By default, H-FISTA does not perform any intrinsic variability correction.

\subsection{Application to dynamic spectra of \psrb}
\label{observed_b}

\begin{figure}
    \centering
    \includegraphics[width=\columnwidth]{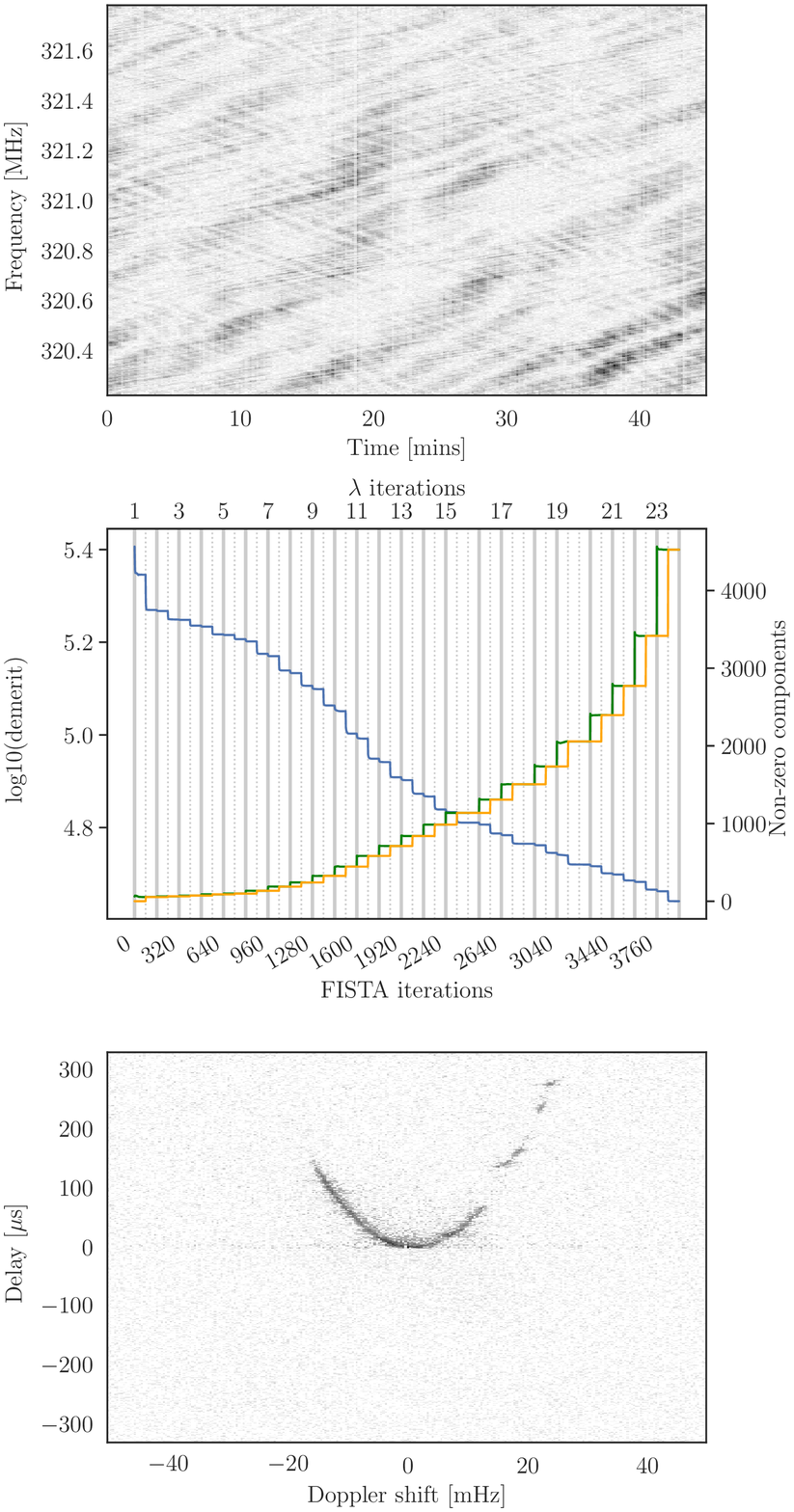}
    \caption{The top panel shows the observed dynamic spectrum of \psrb\ in inverted grey scale after normalisation to remove striation. Middle panel shows the diagnostic plot of H-FISTA, see the caption of Fig. \ref{fig:dynspec_smiley_125_noisefree} for details. The bottom panel is the dense wavefield obtained by H-FISTA (with the transfer function chosen for contrast on significant components of the wavefield).}
    \label{fig:ds_0834}
\end{figure}

We begin with a dynamic spectrum of \psrb, already analysed by \citet{2008MNRAS.388.1214W}. The data were recorded at MJD of $53009$ with the Arecibo Observatory using the Wideband Arecibo Pulsar Processor \citep[WAPP,][]{2000ASPC..202..275D}. The observation was at the centre radio frequency of $321$~MHz with bandwidth of $1.56$~MHz. The data were split into 1024 frequency channels, and a total of $270$ $10$-second sub-integrations. The dynamic spectrum shows striation and thus we normalise it as discussed above. The dynamic spectrum after normalisation is shown in the top panel of Fig.~\ref{fig:ds_0834}.

H-FISTA is stopped by the spatial criterion after $23$ values of $\lambda$. A diagnostic plot is shown in the middle panel of Fig.~\ref{fig:ds_0834}. At the selected step, the sparse wavefield consists of $4526$ non-zero components. The demerit drops rapidly during the first H-FISTA loop before decreasing more gradually during a few steps. This is followed by more rapid decrease during a few steps before more gradual decrease yet again. As seen by the lack of more than one dashed line between any continuous grey lines, very little hard thresholding is taking place.

The single step procedure to obtain a dense solution results in the wavefield presented in the bottom panel of Fig.~\ref{fig:ds_0834}. In this case the phase retrieval with H-FISTA is quite successful. Other than a small amount of power at zero delay, spread across various doppler values\footnote{The power at zero delay is likely a consequence of the simple approach to the removal of striation.}, there are no obvious artefacts present and wavefield at negative delay values is consistent with noise. Comparing to Fig.~3 of \citet{2008MNRAS.388.1214W}, we see our retrieved wavefield is very similar to that earlier result.

\subsection{Application to dynamic spectra of \psra}
\label{observed_a}

\begin{figure}
    \centering
    \includegraphics[width=\columnwidth]{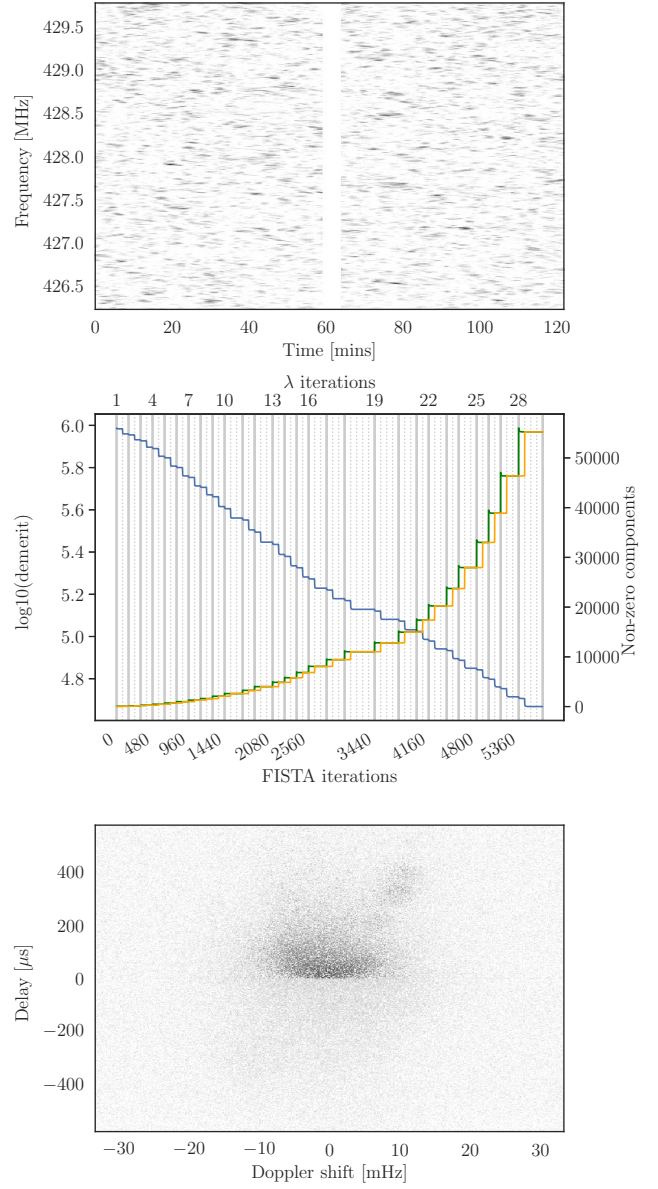}
    \caption{The top panel shows the observed dynamic spectrum of \psra\ in inverted grey scale. The white vertical stripe in the top panel is a gap between observations. The middle panel shows the diagnostic plot of H-FISTA, see the caption of Fig. \ref{fig:dynspec_smiley_125_noisefree} for details. The bottom panel is the dense wavefield obtained by H-FISTA (with the transfer function chosen for contrast on significant components of the wavefield).}
    \label{fig:dynspec_53873}
\end{figure}

We now analyse a dynamic spectrum of \psra\ based on observations reported by \citet{2013ApJ...779...99W}, recorded at the Arecibo Observatory using the Arecibo Signal Processor \citep[ASP,][]{2007PhDT........14D}. The dynamic spectrum, shown in the top panel of Fig.~\ref{fig:dynspec_53873}, corresponds to the observations made on MJD~$53873$, spanning $3.555$~MHz of bandwidth centred at $428$~MHz and split into $4096$ frequency channels. The data consists of $487$ sub-integrations, each 15 seconds long for a total of over $2$ hours. These totals include a gap of $285$ seconds near the middle of the observation.

Before starting the H-FISTA algorithm, we determined that no low pass filtering is required (see \S~\ref{intrinsic_variability}) but we do generate a mask to cover the gap in time around the 60-th minute. With this basic pre-processing, we start H-FISTA with default values of parameters as described in \S~\ref{H_FISTA}. After iterating through $28$ $\lambda$ values, the sparsity-based stopping criterion is triggered; at this stage, the sparse solution consists of $55225$ non-zero wavefield components, or about $3$ per cent of the total phase space. 

The H-FISTA diagnostic diagram, shown in the middle panel of Fig.~\ref{fig:dynspec_53873}, looks qualitatively similar to that for the failed case of locally dense phase retrieval. There is only one episode of hard-thresholding reducing the component count.

Following the procedure described in \S~\ref{densify}, we obtain a dense wavefield from the sparse solution; the result is shown in the bottom panel of Fig.~\ref{fig:dynspec_53873}. The dense solution has a few notable features: the wavefield has a relatively dense core region (near the zero delay and zero doppler-shift) as well as an asymmetric partially-filled arc extending to larger delays. Close inspection of the core reveals that the wavefield appears to be fragmented, that is, multiple large values are interleaved with low values of the wavefield. This behaviour is unsurprising for a $l_1$-regularised optimisation, which tends to select a single parameter from among a group of correlated parameters.

Comparing to the result obtained by \citet{2013ApJ...779...99W}, who used cyclic spectra as input, there are a few easily discerned differences. First of all, the features in the wavefield retrieved in this work are not as sharp as in the previous work. This also results in lower contrast to the noise in the wavefield which is not surprising given we are spreading the same amount of available power across a larger number of components. On the other hand, our approach is not affected by the scattered power visible at low positive delays in Fig.~8 of \citet{2013ApJ...779...99W}.

\section{Discussion}
\label{discussion}

In this section, we comment on the main issues that we are aware of with the method we have presented, and also suggest some possible avenues for improving the performance of the algorithm.

\subsection{Error recognition}
We have attempted to provide H-FISTA with default parameters that should allow it to work without much user input. However, our testing was limited to a small number of input dynamic spectra and users are likely to encounter cases where the results can be improved with some tweaks of the values of those parameters. Unfortunately there are usually no recipes for how the parameters should be adjusted in order to improve the results -- users have to explore the parameter space to some degree -- but an important first step is to recognise problems.

As usual for any in-line hologram, a clear indication of the phase retrieval not succeeding is the presence of twin image components in the wavefield \citep[see e.g.,][]{Liu:87,Guizar-Sicairos:12}. During the H-FISTA iterations, this contamination is seen as the brighter parts of the twin image (i.e. an inverted copy of the wavefield) appearing in one or more areas of the wavefield --- usually in the vicinity of bright regions. This phenomenon is readily understood: the Fourier transform of the dynamic spectrum is a convolution of the wavefield with its twin, and the phase retrieval process can also be thought of as a deconvolution; errors in the phase retrieval thus leave remnants of convolved structure in the wavefield model.

With H-FISTA, which gradually builds up structure in the wavefield model, deconvolution errors can sometimes be spotted at an early stage in the process (i.e. when the regularisation is strong). Naturally, deconvolution errors are easiest to spot when the underlying wavefields are indeed highly sparse, as the features are visually sharper in this case.

The twin image may also manifest itself at a low level in its ``proper place'' -- at negative delays -- in the fully-dense wavefield model. This type of error is relatively straightforward to deal with: it seems to arise when H-FISTA stops too soon, and that can be remedied by relaxing the relevant stopping criterion. A good, dense wavefield solution should appear entirely noise-like in the negative delay half-space. Of course the wavefield may contain many more components if H-FISTA is pursued to smaller values of $\lambda$, whereas the whole approach relies on sparsity to regularise the solution, so this is potentially dangerous. However, our default limit is a sparsity of only $3$ per cent, which is quite conservative, so there is scope for users to relax the stopping criterion if the data demand it. Our experience with high signal-to-noise dynamic spectra is that H-FISTA is actually remarkably stable even with models that have component densities much higher than 3\%. At low signal-to-noise, on the other hand, users might need stricter limits on component numbers.

\subsection{H-FISTA models of locally dense wavefields}
In \S3 we showed the result of applying H-FISTA to two examples of noise-free synthetic dynamic spectra, the first of which provided a perfect reconstruction of the input model whereas the second one failed to find the correct model. In both cases the input model was fairly sparse, but the principal difference between the two cases was the level of sparsity --- the successful case had approximately half the number of components as the case that failed, within an overall wavefield morphology that is otherwise similar. Indeed it appears to be the local density of the wavefield that causes problems for H-FISTA, as we also tested denser input wavefields and they too failed.

When H-FISTA is working on real data there is no clear cut division between success and failure -- the issues are more subtle -- but here too we can see that H-FISTA struggles if the underlying wavefield is dense. The wavefield  retrieved from the dynamic spectrum of \psra\ in \S~\ref{observed_a}, can be seen to be locally dense in the analysis of \citet{2013ApJ...779...99W} (see their Fig.~8). By contrast the wavefield we have obtained, shown in the bottom panel of Fig.~\ref{fig:dynspec_53873}, is fragmented, not smooth, and shows many instances of strong components surrounded by much weaker ones (comparable in strength to the noise). This behaviour is seen everywhere that significant signal is present in our wavefield model, but a particularly important case is the signal at the origin, which is much brighter -- by almost an order of magnitude -- than any other component in the model. 

That concentration of power in the pixel at the origin is at least partly responsible for the diffuse appearance of the faint structure at large delay in the bottom panel of Fig.~\ref{fig:dynspec_53873}, as follows. The Fourier transform of the dynamic spectrum is a sum of products of components drawn pairwise from the wavefield, and much of that structure is contributed by terms in which one of the components is located close to the origin --- because the signals are usually strongest in the vicinity of the origin. Now suppose that the true wavefield has, for example, a feature at large delay with a doppler width $\delta \omega$, and a comparable width to the strong signals around the origin. If our model wavefield is, instead, approximately a delta-function near the origin it follows that the feature at large delay will be forced to have a width $2\,\delta\omega$ in order to match the range of doppler-differences in the data.

It is hardly surprising that H-FISTA tends to produce a fragmented model where the underlying wavefield is actually smooth: in penalising models by their $l_1$ norm we are encouraging exactly that result. Indeed LASSO is known to preferentially select a single variable from a group of correlated variables \citep[see e.g.,][]{ZouHastie05}. As such, H-FISTA is better suited to phase retrieval in instances where the underlying wavefields are sparse not only globally but also locally, such as presented in \S~\ref{observed_b} for the case of \psrb. However, notwithstanding the problems just described, the models produced by H-FISTA may prove adequate even for relatively dense wavefields such as the case of \psra\ --- depending on the specifics of the user's data and the ultimate goal of the analysis.

Finally, we note that even if a spectrum is locally dense, if the adjacent components vary greatly in strength it may appear effectively sparse at each level of regularisation. If so, H-FISTA can be expected to return better results than if all the nearby components have a similar strength.

\subsection{The influence of H-FISTA parameters}
Although our software has default settings for all of the parameters needed for H-FISTA, users can override each of these choices to control the algorithm as desired. We offer the following, relevant commentary.

\subsubsection{Number of FISTA iterations}
If the number of FISTA iterations is too small then phase retrieval is likely to fail simply because the model wavefields at each level of the hierarchy are not sufficiently close to optimal. For example, we found that $N_{\rm iter}=40$ (or less) was typically too small to yield good results on our test data.

While the assertion that having too few iterations can cause problems should not be surprising, a more unexpected result is that increasing $N_{\rm iter}$ can sometimes lead to failure. For example: in one test on the dynamic spectrum analysed in \S~\ref{noise_free_success} we found that increasing $N_{\rm iter}$ from our default value of $80$ continued to give perfect results until  $N_{\rm iter}=160$, which failed. In this example, though, it appears that increasing the number of iterations is not really the cause of the failure, because further increase met with success. Instead we interpret this as a lack of robustness in the algorithm and a simple case of bad luck. However, the majority of parameter combinations enable successful twin image removal and wavefield retrieval.

\subsubsection{Buffer zone of negative delays}
As already discussed in \S~\ref{init_h_support_constraint}, a buffer of allowed negative delays may be necessary to accommodate a non-zero delay in the strongest component of the true wavefield. However, this buffer zone should not cover a large range in delay, otherwise it may be difficult for the algorithm to separate the wavefield from its twin image. We recommend that users do not increase the size of the buffer from the default unless they are sure that the data demand it.

\subsubsection{Scaling of $\lambda$}
In order to hierarchically retrieve more wavefield components, we sequentially lower $\lambda$ as defined in equation~\ref{lambda_sequence}. Increasing $\eta_\lambda$ would result in faster progress through the range of $\lambda$ values of interest, which is desirable. However, if the steps in successive values of $\lambda$ are too large then it may be more difficult to separate components of the wavefield from those of its twin. Our default value of $\eta_\lambda=1.15$ was chosen as a good balance between speed and likelihood of success, for the dynamic spectra that we tested the algorithm on. If the wavefield has a particularly narrow distribution of component moduli then a smaller value of $\eta_\lambda$ may be appropriate.

\subsection{Possible improvements of the method}
No doubt H-FISTA could be improved in a great variety of ways; but in application to real data (\S4) the most objectionable feature of the current algorithm seems to be its tendency to produce fragmented models when the underlying wavefield is actually smooth. Although this behaviour is inherent in the sense that $l_1$ regularisation encourages sparsity, it is possible that a different set of basis functions could yield a better result. For example: a wavelet basis might offer a much better representation of a smooth wavefield, for a given number of non-zero coefficients, than the delta-function basis we have used. It should be a straightforward matter to explore the relative merits of different basis functions, because the gradient of the demerit follows immediately from the basis transformation itself --- as with the Fourier relationship in equation (A15), for example. We note that when considering a different set of basis functions it is appropriate to revisit the choice of the initial wavefield, which in our case was a delta-function at the origin.

A different way of tackling the same problem is to use the so-called ``elastic net'' regularisation \citep{ZouHastie05} instead of LASSO. In brief, the former uses both $l_1$ and $l_2$ regularisation as opposed to the latter which only uses $l_1$. One of the motivations for elastic net was to help address the problem of LASSO preferentially choosing a single variable from a group of related ones, and to perform better in situations with relatively few constraints compared to number of variables.

Finally we note that the difficulty of determining a wavefield model would be lessened considerably if the input dataset contained some explicit phase information. In a dynamic spectrum there is no such information, but in a dynamic cyclic spectrum \citep{2011MNRAS.416.2821D} there is, and from the outset H-FISTA was designed with a view to being easily modified for application to cyclic spectroscopy. The current generation of pulsar instrumentation does not provide cyclic spectra as a standard data product, so for contemporary, or indeed historical datasets\footnote{We note that for much of the historical data the resolution and/or signal-to-noise ratio of the dynamic spectra may not be sufficient for phase retrieval.}, we are obliged to retrieve phases if we want to know the structure of the wavefield. We note that instruments capable of providing cyclic spectra as a standard data product are in development for new generation of wide-band receivers \citep{2021ApJ...913...98D}.

\section{Conclusions}
\label{conclusions}
We have described H-FISTA: a sparse phase retrieval method for pulsar dynamic spectra. This approach uses the $l_1$ norm of the wavefield to regularise fits to the data, with increasingly detailed models built up as the level of regularisation is progressively lowered. We applied this method to both simulated and observed dynamic spectra. With noise-free simulations we demonstrated that perfect reconstruction can be achieved if the wavefield is sufficiently sparse, but this outcome is not guaranteed and the algorithm can fail to converge on the correct model even at quite modest local wavefield densities of order 20 per cent. When applied to real data H-FISTA performed well on a test dynamic spectrum for \psrb,  in which the scattering is highly anisotropic and the wavefield is thus instrinsically very sparse. On denser wavefields, such as exhibited by \psra, the method struggles as it tries to impose sparsity that is not present in the data. Nevertheless, even in this case the model wavefield does not appear to be badly corrupted by the twin image and H-FISTA may prove useful for modelling similar datasets. Our software is freely available and we encourage pulsar astronomers to apply it to their own dynamic spectra.

\section*{Acknowledgements}

The authors are grateful to Willem van Straten and Paul Demorest for fruitful discussion which helped formulate early ideas for this work. We would also like to thank Tim Spenger for their help with identifying a bug in the implementation of the hard thresholding. We are grateful to the anonymous referee for their useful comments.

Data for this project were collected using the facilities of the Arecibo Observatory, a facility of the National Science Foundation operated under cooperative agreement by the University of Central Florida in alliance with Yang Enterprises, Inc. and Universidad Metropolitana.

This research has made use of NASA’s Astrophysics Data System. We also acknowledge the use of \textsc{NumPy} \citep{harris2020array},  \textsc{SciPy} \citep{2020SciPy-NMeth}, \textsc{Astropy} \citep{2022ApJ...935..167A}, and \textsc{matplotlib} \citep{Hunter:2007} libraries.

\section*{Data Availability}

All the dynamic spectra presented in this work are available publicly at this digital object identifier: \href{https://doi.org/10.5281/zenodo.7007226}{10.5281/zenodo.7007226}. The code to perform H-FISTA on these or any other data is available publicly at \href{https://github.com/sosl/H-FISTA/}{https://github.com/sosl/H-FISTA/}. The code repository includes notebooks which were used to produce the results presented in this work.
 


\bibliographystyle{mnras}
\bibliography{HFISTA} 




\appendix

\section{Gradient evaluation}
\label{Wirtinger}
Here we introduce the Wirtinger derivative and derive the corresponding gradient of the demerit, $f$.

\subsection{Wirtinger derivatives}

The Wirtinger derivatives with respect to the variable $z=x+i\,y$ ($z\in\mathbb{C}$; $x,y\in\mathbb{R}$), and its complex conjugate, $z^*$, are:
\be
\frac{\partial\ }{\partial z}\equiv\frac{1}{2}\left(\frac{\partial\ }{\partial x}-i\,\frac{\partial\ }{\partial y}\right),
\qquad {\rm and}\qquad
\frac{\partial\ }{\partial z^*}\equiv\frac{1}{2}\left(\frac{\partial\ }{\partial x}+i\,\frac{\partial\ }{\partial y}\right).
\label{eq:wirtinger_def}
\ee
In cases where the function of interest, $f(z)$, is analytic in $z\in\mathbb{C}$ (i.e. it satisfies the Cauchy-Riemann conditions), the Wirtinger derivatives evaluate to
\be
\frac{\partial f}{\partial z}=f^\prime(z),
\qquad {\rm and}\qquad
\frac{\partial f}{\partial z^*}=0,
\ee
where $f^\prime$ denotes the usual derivative of a complex analytic function. However, the utility of the Wirtinger derivatives is that they are not restricted to analytic functions.

As an example we consider the case
\be
f=|z|^2=z^*z,
\label{eq:square_modulus}
\ee
for which the definitions (\ref{eq:wirtinger_def}) yield the results
\be
\frac{\partial f}{\partial z}=z^*,
\qquad {\rm and}\qquad
\frac{\partial f}{\partial z^*}=z,
\ee
even though $f$ in this case is not analytic, so $f^\prime$ does not exist.

The functional form in equation~\ref{eq:square_modulus} is encountered when undertaking a least-squares fit to data, if the model parameters are complex. This example also illustrates a general rule for the application of Wirtinger derivatives: in practice it is not necessary to use the definitions in equation~\ref{eq:wirtinger_def} to evaluate the derivatives; instead we can simply write the function as $f=f(z,z^*)$ on the domain $\mathbb{C}^2$ and form the partial derivative with respect to $z$ holding $z^*$ constant, and vice versa. At first sight this may be confusing, because when we restrict the domain to $\mathbb{C}$, $z^*$ is fixed by $z$ and vice versa. However, it is an approach that can be rigorously justified and it effects a considerable simplification in the evaluation of derivatives \citep{2009arXiv0906.4835K}.

For functions of several variables, $z_n=x_n+i\,y_n$ ($z_n\in\mathbb{C}$; $x_n,y_n\in\mathbb{R}$; $n=1,2,\dots,N$), we follow  \citet{1983IPCRS.130...11B} and define the gradient as
\be
{\bf\nabla}\ \equiv\left(\frac{\partial\ }{\partial z_1^*},\frac{\partial\ }{\partial z_2^*},\dots,\frac{\partial\ }{\partial z_N^*}\right)^T,
\label{eq:wirtinger_grad}
\ee
where $T$ denotes the transpose. This definition conforms to the expectation that ${\bf\nabla}f$ is the direction of steepest increase of $f$. The first order expansion of a real-valued function, $f({\bf z})$, around a point ${\bf z}_0$ is then
\be
f({\bf z})\simeq f({\bf z}_0)+2\,{\rm Re}\{\langle({\bf\nabla}f)_0^\dagger\,\cdot ({\bf z}-{\bf z}_0)\rangle\}.
\ee
where $\dagger$ denotes the Hermitian conjugate.

\subsection{Gradient of the demerit for fitting to dynamic spectra}

In this paper we are concerned with matching models to measured dynamic spectra, $D(\nu,t)$, and our model spectra take the form
\be
Z(\nu,t)=H(\nu,t)H^*(\nu,t),
\ee
where $H(\nu,t)$ is the electric field as a function of radio-frequency, $\nu$, and time, $t$.
We quantify how well our model fits the data by calculating the demerit
\be
\label{demerit_appendix}
f\equiv\frac{1}{2} \sum R^2,
\ee
in terms of the residual
\be
R = Z(\nu,t) - D(\nu,t),
\ee
and the sum in equation~\ref{demerit_appendix} is taken over the full range of $(\nu,t)$ that we wish to model. We need to evaluate derivatives of the demerit with respect to each of the various model parameters. Note that, because $f$ is real, $\partial f/\partial z^*=(\partial f/\partial z)^*$.

We evaluate the Wirtinger derivatives of the demerit with respect to a set of parameters that characterises $H(\nu,t)$. We introduce the (two-dimensional) Fourier transform $h=\mathcal{F}\left(H\right)$ in discrete form
\be
h_j=\sum\limits_k H_k\,\exp\left\{2\pi i(\tau_j\nu_k+\omega_jt_k)\right\},
\ee
with inverse $H=\mathcal{F}^{-1}\left(h\right)$ in discrete form
\be
H_k=\frac{1}{N_\nu N_t}\sum\limits_j h_j\,\exp\left\{-2\pi i(\tau_j\nu_k+\omega_jt_k)\right\},
\ee
where the indices $j$ and $k$ each range over $N_\nu\times N_t$ distinct values, thus providing a complete description of $H$ and $h$. We refer to the variables $\tau$ and $\omega$ as delay and doppler-shift, respectively, and we refer to $h$ as the wavefield. We choose $(\tau,\omega)$ as our modelling space, and we require the model wavefield to be causal, i.e. $h(\tau,\omega)=0\;\;\forall\;\;\tau<0$.

As per equation~\ref{eq:wirtinger_grad}, the appropriate gradient operator is
\be
{\bf\nabla}\ =\left(\frac{\partial\ }{\partial h_1^*},\frac{\partial\ }{\partial h_2^*},\dots,\frac{\partial\ }{\partial h_N^*}\right)^T. 
\ee
And we can evaluate each derivative by making use of the chain rule:
\begin{eqnarray}
\frac{\partial f}{\partial h_j^*}=&\mathlarger{\sum}\limits_k \left\{ \frac{\partial f}{\partial H_k^*}\frac{\partial H_k^*}{\partial h_j^*}    +   \frac{\partial f}{\partial H_k}\frac{\partial H_k}{\partial h_j^*} \right\}\hskip 3.2cm\\
=&\mathlarger{\sum}\limits_k \frac{\partial f}{\partial H_k^*}\frac{\partial H_k^*}{\partial h_j^*}\hskip 5.0cm\\
=&\frac{1}{N_\nu N_t}\mathlarger{\sum}\limits_k \frac{\partial f}{\partial H_k^*}\,\exp\left\{2\pi i(\tau_j\nu_k+\omega_jt_k)\right\},\hskip1.5cm
\end{eqnarray}
Thus, up to a normalising constant, the gradient with respect to $h$ is just the Fourier transform of the gradient with respect to $H$, and the components of the latter evaluate to
\be
\frac{\partial f}{\partial H_k^*}\;=\; 
R(\nu_k,t_k)\,H(\nu_k,t_k)\;=\; R_k\,H_k\,.
\ee

\section{Estimating the Lipschitz constant}
\label{Lipschitz_calculation}
As noted earlier, we have been unable to derive an analytic estimate of the Lipschitz constant appropriate to our context, but the backtracking procedure described in \S\ref{backtracking} mitigates that difficulty. To proceed we need only an approximate initial estimate of $L$, which can then be increased as necessary to satisfy the condition (\ref{backtracking_eq}); a suitable estimate can be made as follows.

Our initial wavefield model is a plane wave at the origin, with the phase set to zero. Now consider the hypothetical case where the dynamic spectrum is simply a constant, so that $D(\nu,t)=\bar{D}\;\;\forall\;\; \nu,t$. A model dynamic field $Y(\nu,t)=\sqrt{\bar{D}}$ will precisely match that dynamic spectrum, so $f(y)=0$ and $\nabla f(y)=0$. We can explicitly evaluate the demerit for any other model $X=Y+\Delta$, from the definition (A8), and in the simple case where $\Delta$ is also a constant, and small compared to $Y$, the result is
\be
f(x)=2\,N_\nu N_t\, \bar{D} \, \Delta^2,
\ee
to lowest order in $\Delta$. The demerit for this model should also satisfy equation~\ref{backtracking_eq}, which simplifies to
\be
f(x)\le \frac{L}{2}|| x-y ||^2.
\ee
Now for constant $\Delta$ we have $x-y=\mathcal{F}(\Delta)$, which is zero everywhere except for a component of amplitude $N_\nu N_t \,\Delta$ at the origin. The backtracking condition thus gives us a lower limit on the Lipschitz constant of
\be
L\ge\frac{4\,\bar{D}}{N_\nu N_t},
\ee
and we use that limit as our initial estimate of $L$.

\section{Computational performance}
\label{performance}

In table~\ref{table:performance}, we present a summary of the performance of the H-FISTA algorithm as described in this article. We present the time required to compute the sparse and dense wavefields for all the cases we presented. The reported times are based on running the algorithm on the author's laptop which has an Intel\textsuperscript{\textregistered} Core\textsuperscript{\texttrademark} i7-1068NG7 processor. We used two threads for all the Fourier transforms.

\begin{table}
\begin{center}
\begin{tabular}{ccccc}
Model & Size & Sparse [min] & Iterations & Dense [min] \\
\hline
NF $12.5$ per cent & $262,144$ & $19.8$ & $11,120$ & N/A \\
$12.5$ per cent & $262,144$ & $11.6$ & $6,880$ & $1.4$ \\
NF $25$ per cent & $262,144$ & $10.0$ & $5,840$ & N/A \\
\hline
\psrb & $276,480$ & $7.8$ & $4,080$ & $1.5$ \\
\psra & $1,994,752$ & $104.2$ & $6,000$ & $16.5$ \\
\hline
\end{tabular}
\caption{Estimated computation time for calculating sparse and dense wavefields, for all the cases presented in this article. ``NF'' indicates noise-free dynamic spectra. We also present the size of the model, that is the total count of pixels in the dynamic spectrum. We note that the iterations count includes the iterations which led to the triggering of one of the stopping criterion and thus are higher than the numbers shown on most of the figures.}

\label{table:performance}
\end{center}
\end{table}

One of the computationally expensive aspects of the algorithm, and one which does not happen at a predictable frequency, is the backtracking. This is one of the main reason for the variation of the algorithm performance. However, we can roughly estimate that H-FISTA takes about one second to run five FISTA iterations per every half a million pixels in the dynamic spectrum on our computational setup.

\bsp	
\label{lastpage}
\end{document}